# When AI Writes Back: Ethical Considerations by Physicians on AI-Drafted Patient Message Replies

Di Hu, MS[1], Yawen Guo, MISM[1], Ha Na Cho, MS[1], Emilie Chow, MD[1], Dana B. Mukamel, PhD[1], Dara Sorkin, PhD[1], Andrew Reikes, MD[1], Danielle Perret, MD[1], Deepti Pandita, MD[1], Kai Zheng, PhD[1]
[1]University of California, Irvine, Irvine, CA, USA

**Abstract**
*The increasing burden of responding to large volumes of patient messages has become a key factor contributing to physician burnout. Generative AI (GenAI) shows great promise to alleviate this burden by automatically drafting patient message replies. The ethical implications of this use have however not been fully explored. To address this knowledge gap, we conducted a semi-structured interview study with 21 physicians who participated in a GenAI pilot program. We found that notable ethical considerations expressed by the physician participants included human oversight as ethical safeguard, transparency and patient consent of AI use, patient misunderstanding of AI's role, and patient privacy and data security as prerequisites. Additionally, our findings suggest that the physicians believe the ethical responsibility of using GenAI in this context primarily lies with users, not with the technology. These findings may provide useful insights into guiding the future implementation of GenAI in clinical practice.*

**Introduction**

Managing large volumes of "in-basket" messages in electronic health record (EHR) systems demands a substantial amount of time and effort, often extending beyond regular clinical hours into what is commonly referred to as "work-outside-work"[1,2]. It constitutes a significant portion of clinicians' workload, contributing to work dissatisfaction, increased stress, and ultimately, burnout[3–5]. Among such in-basket messages, patient-initiated messages submitted through the patient portal account for the majority of processing work time[2]. While patient–provider messaging has improved patient outcomes and strengthened patient–provider relationships[6,7], it has also introduced challenges that strain clinician efficiency and well-being. These challenges intensified since the COVID-19 pandemic, when the volume of patient messages nearly doubled compared to pre-pandemic levels, further increasing in-basket workload and exacerbating clinician burnout[8]. Addressing the in-basket burden has thus become a pressing issue, prompting the exploration of novel solutions such as artificial intelligence (AI)-driven clinical efficiency improvement tools[9].

In the recent few years, generative AI (GenAI) has been explored as a promising tool to facilitate the processing of patient messages in various clinical settings. Early investigations into GenAI-drafted replies have primarily focused on evaluating their impact on in-basket work time, message length, and physician burden, as well as assessing their accuracy, empathy, readability, and usefulness[10–13]. Findings generally reported positive feedback, with physicians perceiving GenAI's potential to enhance patient–provider communication, reduce time burden, and alleviate workload[10–15]. Some studies also found that AI-generated drafts were comparable to, or even preferred over, physician-written responses, particularly in addressing empathy[14,15]. Despite these benefits, ethical considerations surrounding the use of GenAI for drafting patient replies in real-world clinical practice remain largely unexplored. These considerations are critical to the successful adoption of GenAI, as they directly impact trust, user acceptance, and the perceived safety and effectiveness of these tools[16–18]. Given that physicians are among the primary users of GenAI for replying to patient messages, their perspectives play a key role in ensuring the tool's ethical and responsible implementation.

To address this knowledge gap, we conducted an interview study to explore physicians' perceptions of and experiences with using a GenAI tool that drafts replies to patient messages, with a particular focus on their ethical considerations. Our study objective was twofold:

1. To assess physicians' ethical concerns regarding the use of GenAI in drafting replies to patient messages.
2. To gather physicians' insights on improving the implementation of GenAI to mitigate potential ethical concerns.

By capturing physicians' ethical considerations on AI-assisted patient communication, this study aims to provide actionable insights for the responsible deployment of GenAI in patient messaging in various clinical settings. The findings may inform AI design improvements, ethical safeguards, and policy recommendations to support sustainable and effective adoption of AI-driven in-basket management.

## Methods

*Recruitment*

Participants of the study were recruited from an academic health center between April and July 2024. Eligibility was limited to clinicians enrolled in a pilot program evaluating the efficacy of MyChart In-Basket Augmented Response Technology (Epic Systems Corporation, Verona, WI, USA), a GenAI tool for automatically drafting responses for clinicians replying to patient messages. At the time of the study, the pilot included only physicians and no other clinicians. This tool integrates into Epic's EHR system and leverages OpenAI's large language model (OpenAI, San Francisco, CA, USA)[19]. This pilot program began in July 2023 with a few early adopters and then expanded incrementally by departments. Family Medicine was added in September 2023, followed by Primary Care in early December, Gastroenterology in late December, and Urology in January 2024. By May 2024, the program was opened to all physicians, with the GenAI draft appearing after they completed an acknowledgment process. Recruitment emails were sent to all physicians in the pilot, and those interested in the study either replied to the email or directly contacted the research team. To ensure sufficient experience with the tool, physicians who had been enrolled for less than three months were excluded. Participation was entirely voluntary, and no compensation was provided. This study was deemed self-exempt by the Institutional Review Board of the University of California, Irvine.

*Participants*

A total of 21 physicians participated in the study. The majority were female (n=15, 71.4%), and the sample consisted of a range of age groups, with the largest proportion (n=8, 38.1%) in the 40–49 age range. Clinical experience was relatively balanced, with similar representation across those with 1–10 years (n=5, 23.8%), 10–20 years (n=8, 38.1%), and over 20 years (n=8, 38.1%) of post-medical school practice. Most participants specialized in either internal medicine or family medicine. More than half of them (n=13, 62.0%) were board-certified in two or more medical specialties, indicating their diverse clinical expertise and interdisciplinary practice. The presence of physicians from different specialties and experience levels allowed for a broader perspective on clinical practice and technology use. At the time of their interview, most participants (n=16, 76.2%) had been enrolled in the pilot program using the GenAI tool to draft patient message replies for 3 to 6 months. Table 1 reports the general characteristics of the sample.

**Table 1.** Participant Characteristics.

| Characteristic | n (%) |
| --- | --- |
| **Gender** | |
| Female | 15 (71.4) |
| Male | 6 (28.6) |
| **Age Group (Years)** | |
| 30–39 | 7 (33.3) |
| 40–49 | 8 (38.1) |
| 50–59 | 4 (19.0) |
| 60+ | 2 (9.5) |
| **Years in Clinical Practice (Post-Medical School)** | |
| 1–10 | 5 (23.8) |

| 10–20 | 8 (38.1) |
|---|---|
| 20+ | 8 (38.1) |
| **Board-Certified Specialties (Physicians May Hold Multiple Certifications)** | |
| Internal Medicine | 8 (38.1) |
| Family Medicine | 6 (28.6) |
| Geriatric Medicine | 3 (14.3) |
| Gastroenterology | 3 (14.3) |
| Psychiatry | 2 (9.5) |
| Urology | 2 (9.5) |
| Physical Medicine & Rehabilitation | 2 (9.5) |
| Other (General Pediatrics, Endocrinology & Metabolism, Sports Medicine, Pain Medicine, Bariatric Medicine, Urogynecology) | 6 (28.6) |
| **Duration of Pilot Participation (Months)** | |
| 3–6 | 16 (76.2) |
| 7–11 | 5 (23.8) |

*Data Collection*
The research team developed a semi-structured interview protocol through multiple rounds of discussion, drawing on their expertise in AI, informatics, and clinical practice. The protocol was revised and finalized after pretesting with one physician representative. The interview questions were organized into three main sections:

1. Physicians' general understanding of GenAI and its capabilities.
2. Physicians' typical workflow when using the piloted GenAI tool to draft replies to patient messages.
3. Physicians' perceived concerns about the tool and their envision for its future application in clinical practice.

The analysis of this study specifically focused on **the third section**, understanding physicians' perspectives on relevant ethical issues and considerations. One example interview question related to this focus was: "*Do you think there are any ethical issues associated with using the AI tool?*" Throughout the interviews, probes and follow-up questions were used as needed to encourage participants to elaborate on and clarify their responses. All interviews were conducted via Zoom, audio-recorded, and transcribed. Each session lasted approximately 30 minutes, and verbal consent for participation and recording was obtained at the beginning of each interview. No identifying information was collected as part of the study.

*Data Analysis*
We applied thematic analysis to analyze the interview transcripts[20]. The first and second authors (DH and YG) familiarized themselves with the transcripts and independently open-coded five transcripts to identify the main patterns in physicians' perspectives and experiences related to the use of GenAI in drafting patient message replies. They compared their initial codes and discussed key takeaways with the research team. Through these discussions, we decided to focus our analysis on understanding physicians' ethical perspectives regarding the use of GenAI-drafted replies. Building on these insights, the first author developed a preliminary codebook centered on ethical considerations in AI-assisted patient messaging. The first author reviewed the codebook with the research team, then applied it to the remaining transcripts, adjusting the codes as new patterns emerged. The research team met regularly to discuss and refine the codes and themes, ensuring that the analysis remained grounded in participant perspectives. Qualitative coding was conducted using ATLAS.ti (ATLAS.ti GmbH, Berlin, Germany).

**Results**
Through thematic analysis, five key themes emerged from our interview data, each highlighting a distinct ethical dimension of physicians' perspectives regarding the use of GenAI in patient messaging. These themes reflect physicians' considerations about maintaining human oversight, ensuring transparency and patient consent, addressing potential misunderstandings of AI's role, safeguarding patient privacy and data security, and clarifying ethical responsibility in AI-assisted communication. These ethical considerations shaped how physicians perceived the tool's implementation and the safeguards necessary for its responsible use. Table 2 provides a summary of these themes, and each is described in detail below with participants' quotes.

**Table 2.** Theme Descriptions.

| Theme | Description |
| --- | --- |
| Human Oversight as Ethical Safeguard | Physician review of AI-generated draft replies ensures ethical standards, accuracy, and accountability of the information communicated. |
| Transparency and Patient Consent of AI Use | Transparency and patient consent are key to addressing ethical concerns in AI-assisted communication. |
| Patient Misunderstanding of AI's Role | Misunderstanding of AI's role could lead to patient distrust or dissatisfaction. |
| Patient Privacy and Data Security as Prerequisites | AI tools must securely store patient data and comply with privacy protection requirements to avoid ethical risks. |
| Ethical Responsibility Lies with the User, Not AI | Ethical use of AI depends on users, emphasizing that physicians need to supervise and thoughtfully apply AI in patient care. |

*Human Oversight as Ethical Safeguard*
Physician participants in this study overwhelmingly agreed human oversight is essential for ensuring ethical and responsible use of GenAI in drafting patient message replies. They viewed GenAI as a tool with potential to assist message reply but emphasized that physicians need to review, edit as needed, and sign off on messages before sending them to patients. Most participants expressed no ethical concerns regarding GenAI-drafted replies as long as this oversight process remained in place and they held the "*ultimate power*" over message content (P2, P4, P6–8, P10, P14, P17–21). As P21 explained:

> *I'm open to that [using GenAI for drafting replies]. As long as I have control over what is going to be sent in my name. It's practically sent in my name, but it's not necessarily written by me. For me. I think it's fully ethical.*

P8 saw GenAI use under oversight as analogous to how they supervise other clinical communications, such as messages written by nurses or trainees, and elaborated "*In that respect it's no different than a reply verbally to a nurse or a reply that a resident or somebody else gives that the physician signs off on.*" The key factor was that the final message must accurately reflect the physician's judgment and intent, aligning with what physicians would write on their own (P2, P4, P7, P19). Beyond ethical concerns, human oversight was also viewed as critical for liability and patient safety. P6 emphasized that ensuring physician review prevents AI from becoming an independent decision-maker in providing medical advice, stating, "*It's not just the machine answering…, There should always be oversight by us*". Several participants explicitly opposed the idea of fully automated AI replying, warning that removing human oversight could lead to not only ethical concerns but also patient safety issues (P2, P19–20). P20 strongly cautioned against autonomous AI:

> *If we count on it [GenAI] solely, like a hundred percent, I think there will be [issues]… If we use it as a pure tool without a physician even reviewing the medical content, it will create ethical and even care, like medical care issues.*

This strong consensus underscores that while physicians see value in AI-assisted messaging, they also recognize that it must remain an assistant, not a substitute for answering patient questions.

### *Transparency and Patient Consent of AI Use*

Another commonly agreed-upon strategy for mitigating ethical concerns is being transparent about GenAI use in replying to patient messages. Some participants believed that providing disclaimers alongside AI-assisted replies was sufficient to address ethical risks (P6, P15, P19). As P15 stated, "*As long as there is the disclaimer, I think the way the world is moving [now]… that should be fine.*" However, others argued that passive disclosure was not enough and that patients should actively consent to the use of GenAI in replying to their messages (P5, P7, P13). P13 suggested that patients should be given a choice:

> *We have to let the patients know that we may respond using AI assisted tools, right? I think they have to know that, if they're not okay with that, then that is an issue…like the patients have to say that they consent to it. I think they have to be okay with it.*

There were also concerns that some patient groups, particularly older adults and those unfamiliar with AI, might be less receptive to GenAI-assisted messaging (P5, P7, P13). P5, a geriatrician, explained:

> *I think, especially in geriatrics, patients are looking more for the human side of medicine. I'm not sure if other fields of medicine might be more, I guess, less of a problem. But in geriatrics, I just think It does matter. [They prefer] a human to talk to.*

A psychiatrist (P7) adjusted their use of GenAI based on their perception of patient attitudes, avoiding GenAI drafts or removing disclaimers for patients with "*technology paranoia*". They explained that this selective approach was intended to prevent unnecessary anxiety about messages that had already been reviewed and approved. Overall, while transparency was considered important for upholding ethical GenAI use, participants differed in their views on the best approach. Some advocated for full patient consent, while others believed a disclaimer alone was sufficient. A few also felt that disclaimers could be removed in special cases to avoid raising unnecessary concerns among patients.

### *Patient Misunderstanding of AI's Role*

While many physicians supported transparency, some worried that patients might misinterpret AI disclaimers, leading to unintended consequences (P1, P7, P16, P19). P7 raised a concern that patients could mistakenly believe GenAI-drafted responses were composed by fully autonomous bots, which might damage trust in patient–physician communication:

> *Even though it says like this [the reply] is either fully or in part [generated] with ChatGPT in the disclaimer. I could see the patient thinking, this is an automated response that hasn't really gone through me. I think that seeing that at the bottom [disclaimer] would sometimes immediately make them doubt anything in the message before.*

P16 further noted that patients' perceptions of AI involvement could influence their satisfaction with care, particularly when billing is involved for patient–physician messaging. The participant explained:

> *If the patient sees that it was actually drafted by AI, but they're having to pay for that time. They might be unhappy with that, but they might not understand that even if it was drafted by AI, it still required a human to verify and send it. So I can see sort of patience being unhappy with being billed when they think it's an AI responding. But, in fact, there's a real person regulating the response.*

These concerns suggest that while transparency is important, the disclaimer or the approach for informing patients about the GenAI used in messaging must be designed carefully to avoid creating confusion or mistrust among patients for this communication.

### *Patient Privacy and Data Security as Prerequisites*

Our participants also considered data security and patient privacy protection to be essential conditions for the ethical use of GenAI in drafting replies. Many participants did not perceive ethical concerns related to patient privacy and data security, given that the AI tool they were piloting was integrated into the EHR and compliant with Health Insurance Portability and Accountability Act regulations[21] (P3, P4, P12, P14). P3 further clarified that they did not perceive the tool as introducing any additional data security risks:

> *I mean I'm not a computer scientist, but I don't think it's any less or more susceptible to data breach than the EMR system that we're currently using. So I don't see an ethical reason for not to use it.*

Similarly, P4 did not foresee additional privacy risks, stating "*As far as an ethical consideration in terms of patient privacy, there's complete patient privacy. It's not giving out any details that I wouldn't already be privy to.*" However, some participants still raised hypothetical concerns about how AI tools access patient data. P14 questioned whether the AI might collect patient information when integrated with the EHR system, potentially resulting in ethical concerns:

> *I think potentially, if there's any security issue. If this AI tool is contained within the EMR, then I don't know if it would be collecting patient information in any sort of way across different charts. That might become an ethical issue.*

The underlying condition raised by physicians in this theme was that strong protections for patient data and privacy are required for ethical GenAI implementation.

### *Ethical Responsibility Lies with the User, Not AI*
The final theme focused on attributing ethical responsibility to the user rather than the GenAI tool itself. Many participants saw no inherent ethical problem with using GenAI to draft replies, arguing that responsibility lies in how clinicians use the tool (P8, P11–12, P21). P8 noted, "*If a physician is not thoroughly vetting the message before he or she sends it. Then there's a huge ethical problem, but the ethics is on the part of the physician, not the AI.*" P21 also summarized this perspective:

> *I don't think it is unethical. I think it [GenAI] just needs to be appropriately used. I think it depends how you're using it. You'll use and make it unethical. I think it's up to the provider or person who is going to use it.*

Concerns were expressed that overworked healthcare staff might become overly reliant on AI, which could lead to ethical issues. P11 warned about the risk of sending "click-through" responses, saying "*For some MAs [medical assistants], because they're overworked.… they may not look into anything. They may just point and click, then there will be a lot of inappropriate messages that get sent to patients.*" Additionally, P12 cautioned that ethical and medical risks could emerge if the physician is not being responsible for their replies, noting that these issues exist even without AI involvement:

> *The biggest question is if the physician is responsible for the actual reply, taking the time to actually think about the response, and giving a thoughtful response to the patient, so that ethical question exists with or without the presence of the AI generated responses.*

Overall, physician participants did not attribute ethical issues to GenAI but emphasized that responsible use depends on clinician engagement. They also highlighted the importance of maintaining human supervision of AI and remaining committed to validated and thoughtful patient care.

### Discussion
The findings of this study indicate that physician participants expressed limited ethical concerns regarding their current use of GenAI in drafting replies to patient messages. However, this result is contingent upon two key conditions: (1) the GenAI tool being embedded within the EHR, minimizing additional data security and patient privacy risks, and (2) the requirement for physicians to review and edit AI-generated drafts to ensure accuracy, accountability, and compliance with ethical and professional standards. These conditions also appear to be essential for increasing AI trustworthiness in a broader healthcare context[16]. Additionally, rather than viewing the AI tool as inherently risky, physicians believed that ethical and medical risks are more likely to stem from irresponsible use rather than the technology itself. This perspective reinforces a broader shift from seeing AI as an independent ethical actor to recognizing human responsibility in its clinical applications[17]. Research on responsible AI in healthcare further underscores that accountability extends beyond users like physicians, as AI-related risks are also shaped by how the technology is developed, integrated, and governed[16,22]. This highlights the role of multiple stakeholders, including application developers, algorithm designers, data providers, and policy regulators, in ensuring AI's ethical use in clinical care[16,23].

Although our physician participants expressed different opinions on whether, when, and how patients should be notified about AI use, they consistently recognized that patients play an important role in determining the ethical acceptability of AI-generated message replies. This emphasizes the need for future research to explore how patients perceive AI-generated replies, their preferences for disclosure, and their concerns about how AI is applied in medical communication, as also recognized in other studies[10,11]. In addition, our findings intersect with the recent discussions on billing policies for electronic patient communications[24]. As billing for patient messages becomes a standard practice, questions arise regarding how AI-assisted replies should be categorized particularly when AI contributes to message composition but still requires clinician oversight and modification. Some participants in our study raised concerns that patients may not fully understand the human role in AI-assisted replies, which could affect their expectations regarding billing and service quality. Future studies should explore how billing policies assess this complicated health delivery method with human and AI collaboration and reevaluate the ethical considerations on billing AI-assisted messages.

Our findings reveal several important physician-informed implications for the ethical implementation of GenAI in patient messaging. Ensuring strong human oversight remains essential, as GenAI tools should be designed to require clinician review and formally signed approval before messages are sent, ensuring accountability and mitigating liability concerns. At the same time, transparent disclosure strategies must be developed to inform patients about AI involvement in a way that avoids misinterpretation or distrust. For patients who may be skeptical of AI-assisted communication, providing opt-in consent options could help build trust and ensure AI integration aligns with patient preferences. Education programs should also be provided to help patients understand how GenAI may be used in their care, how it can alleviate clinician burnout, and how clinicians remain actively involved in overseeing and personalizing their care despite AI integration. Increasing awareness that GenAI tools can reduce clinician burnout while enabling accurate and more empathetic communication may lead to greater patient acceptance and reduce corresponding ethical concerns[11,25]. Additionally, maintaining rigorous data security standards is critical, ensuring GenAI tools operate within established EHR infrastructure with protections on messages storage and access to minimize external data risks[26]. Finally, clinician training on both the benefits and risks of GenAI, including potential biases and over-reliance, can support responsible adoption and prepare clinicians to navigate ethical considerations effectively. These strategies can help ensure that AI-assisted messaging enhances clinical workflows while maintaining transparency, professional ethics, and patient trust.

This study has several limitations. First, due to our broader goal of understanding clinician experiences with using GenAI to draft replies to patient messages, we only recruited and interviewed physicians, as other healthcare providers were not included in the GenAI pilot at the time of the study. While physicians' perspectives are valuable as primary users of GenAI-drafted replies, they may not fully represent those of other providers, such as nurse practitioners and physician assistants. With the recent expansion of the pilot to include a wider range of providers, our next step will be to explore their perspectives to further understand diverse ethical considerations in a more applied clinical context. Second, our sample size was relatively small and consisted of self-selected participants who opted into the GenAI pilot, which might introduce bias toward individuals already receptive to AI use in clinical settings. Future research should address these limitations by drawing random samples from a larger user group and including diverse participants in different clinical roles beyond early adopters. As the tool becomes more widely available, it will also be important to conduct longitudinal studies to examine how physician attitudes toward AI-generated messaging change over time, particularly as AI capabilities improve and institutional policies adapt. Additionally, future studies should explore how user demographics, roles, and clinical experiences shape ethical considerations toward this AI tool, as well as how users' attitudes and perceptions ultimately influence their use of the technology in practice.

**Conclusion**
Unlike prior studies that primarily focus on assessing the efficacy and efficiency of GenAI in drafting patient message replies, this study identifies key ethical concerns and the conditions under which physicians perceive GenAI as acceptable for this purpose. The findings highlight the importance of human oversight, transparency, data security, patient privacy, and patient understanding in ensuring the ethical use of GenAI for patient messaging. Recognizing that ethical risks arise from how GenAI is developed, integrated, and used rather than from the technology itself, the study underscores the need for robust governance and user training to foster responsible GenAI adoption in clinical practice. These insights may help inform healthcare administrators, AI developers, and clinicians as they work to guide the ethical and responsible implementation of GenAI in patient care.

**Funding and Acknowledgement**
This study was funded in part by the NIH/National Center for Advancing Translational Sciences (NCATS) grant number 1UM1TR004927. The authors acknowledge all physician participants for their valuable time and contributions.